\documentclass[10pt]{article}

\usepackage{amsmath,amssymb,amsthm}
\usepackage{amsxtra}
\usepackage{amsfonts}
\usepackage{amssymb}
\usepackage{url}
\usepackage{hyperref}
\usepackage{esint}
\usepackage{eucal}
\usepackage{mathrsfs}
\usepackage{cite}
\usepackage{graphicx,graphics}
\usepackage{stackengine}
\setstackgap{S}{1pt}

\newcommand\labsect[1] {\label{sect:#1}}

\newcommand\eq[1] {(\ref{#1})}
\newcommand{\bfm}[1]{\mbox{\boldmath ${#1}$}}

\newcommand{\nonum}{\nonumber \\}
\newcommand{\beqa}{\begin{eqnarray}}
\newcommand{\eeqa}[1]{\label{#1}\end{eqnarray}}
\newcommand{\beq}{\begin{equation}}
\newcommand{\eeq}[1]{\label{#1}\end{equation}}
\newcommand{\Grad}{\nabla}
\newcommand{\Div}{\nabla \cdot}
\newcommand{\Curl}{\nabla \times}

\newcommand{\Md}{\partial}

\newcommand{\Ga}{\alpha}
\newcommand{\Gb}{\beta}
\newcommand{\Gd}{\delta}
\newcommand{\Ge}{\epsilon}
\newcommand{\Gve}{\varepsilon}
\newcommand{\Gf}{\phi}

\newcommand{\Gg}{\gamma}

\newcommand{\Gk}{\kappa}

\newcommand{\Gl}{\lambda}
\newcommand{\Gn}{\eta}
\newcommand{\Gm}{\mu}
\newcommand{\Gv}{\nu}

\newcommand{\Gt}{\theta}

\newcommand{\Gr}{\rho}

\newcommand{\Gs}{\sigma}

\newcommand{\Gj}{\tau}

\newcommand{\Go}{\omega}
\newcommand{\Gx}{\xi}
\newcommand{\Gy}{\psi}
\newcommand{\Gz}{\zeta}

\newcommand{\GF}{\Phi}

\newcommand{\BGb}{\bfm\beta}

\newcommand{\BGe}{\bfm\epsilon}
\newcommand{\BGve}{\bfm\varepsilon}

\newcommand{\BGn}{\bfm\eta}
\newcommand{\BGm}{\bfm\mu}

\newcommand{\BGr}{\bfm\rho}

\newcommand{\BGs}{\bfm\sigma}

\newcommand{\BGF}{\bfm\Phi}
\newcommand{\BGG}{\bfm\Gamma}
\newcommand{\BGL}{\bfm\Lambda}

\newcommand{\BGY}{\bfm\Psi}

\newcommand{\CE}{{\cal E}}

\newcommand{\CJ}{{\cal J}}

\newcommand{\CT}{{\cal T}}

\newcommand{\BCA}{{\bfm{\cal A}}}

\newcommand{\BCC}{{\bfm{\cal C}}}
\newcommand{\BCD}{{\bfm{\cal D}}}

\newcommand{\BCI}{{\bfm{\cal I}}}

\newcommand{\bpm}{\begin{pmatrix}}
\newcommand{\epm}{\end{pmatrix}}

\def\ii{{\rm i}}

\def\Bb{{\bf b}}

\def\Bd{{\bf d}}
\def\Be{{\bf e}}
\def\Bf{{\bf f}}
\def\Bg{{\bf g}}
\def\Bh{{\bf h}}

\def\Bj{{\bf j}}
\def\Bk{{\bf k}}

\def\Bm{{\bf m}}
\def\Bn{{\bf n}}

\def\Bp{{\bf p}}
\def\Bq{{\bf q}}
\def\Br{{\bf r}}
\def\Bs{{\bf s}}
\def\Bt{{\bf t}}
\def\Bu{{\bf u}}
\def\Bv{{\bf v}}
\def\Bw{{\bf w}}
\def\Bx{{\bf x}}

\def\BA{{\bf A}}

\def\BC{{\bf C}}
\def\BD{{\bf D}}
\def\BE{{\bf E}}
\def\BF{{\bf F}}
\def\BG{{\bf G}}

\def\BI{{\bf I}}
\def\BJ{{\bf J}}
\def\BK{{\bf K}}
\def\BL{{\bf L}}
\def\BM{{\bf M}}
\def\BN{{\bf N}} 

\def\BP{{\bf P}}
\def\BQ{{\bf Q}}
\def\BR{{\bf R}}
\def\BS{{\bf S}}
\def\BT{{\bf T}}

\def\BV{{\bf V}}
\def\BW{{\bf W}}

\def\BY{{\bf Y}}

\usepackage{graphics}
\usepackage{amssymb}

\title{A unifying perspective on linear continuum equations prevalent in science. Part III: Canonical forms for dynamic
equations with moduli that may, or may not, vary with time}
\author{}
\date{}
\begin{document}
\maketitle
\vskip -.5cm
\centerline{\large Graeme W. Milton}
\centerline{Department of Mathematics, University of Utah, USA -- milton@math.utah.edu.}
\vskip 1.cm
\begin{abstract}
  Enlarging on Parts I and II  we write more equations in the desired format of the extended abstract theory of composites.
  We focus on a multitude of full dynamic equations, including equations where the medium is moving or otherwise changing in time. The motivation is that results and methods in the theory of
  composites then extend to these equations.
\end{abstract}
%%%%%%%%%%%%%%%%%%%%%%%%%%%%%%%%%%%%%%%%%%%%%%%%%%%%%%%%%%%%%%%%%%%%%%%% 
\section{Introduction}
\setcounter{equation}{0}
\labsect{1}
%%%%%%%%%%%%%%%%%%%%%%%%%%%%%%%%%%%%%%%%%%%%%%%%%%%%%%%%%%%%%%%%%%%%%%%%%%%%%%%%%%%%%%%%%%%%%%%%%%%%%%%%%%%
In this Part  III  we continue casting a multitude of linear science equations in
the form encountered in the extended abstract theory of composites:
\beq \BJ(\Bx,t)=\BL(\Bx,t)\BE(\Bx,t)-\Bs(\Bx,t),\quad \BGG_1\BE=\BE,\quad\BGG_1\BJ=0,
\eeq{ad1}
where we now allow the fields and moduli to depend on $\Bx$ and possibly $t$. 
%The first equation in \eq{ad1}, i.e., the constitutive law, is typically taken to be local in spacetime if we
%allow $x_4$ to represent time), i.e. $\BJ(\Bx,t)=\BL(\Bx,t)\BE(\Bx,t)-\Bs(\Bx,t)$.
The field $\Bs(\Bx,t)$ is the source term. Letting $(\Bk, \Go)$ be the Fourier
coordinates associated with coordinates $(\Bx, t)$ in real space, $\BGG_1(\Bk,\Go)$ is a projection operator in Fourier space, while
$\BL(\Bx,t)$ acts locally in real spacetime. Going beyond the static, quasistatic, and time harmonic equations that we
had expressed in this form in Parts  I  and II \cite{Milton:2020:UPLI, Milton:2020:UPLII} our focus in this Part III
is on dynamic wave equations, both with material properties that do not vary with time, or which vary with time due to motion of the underlying medium
or due to some intrinsic change of material properties. In Part IV
\cite{Milton:2020:UPLIV} we
will consider equations that involve higher order gradients of the fields.

The wave equations that we consider here are ones where there are no viscous (lossy) terms. If such terms are present then this often indicates
one should have a convolution in time. To incorporate such viscous terms or convolutions in time one can separate our source $\Bs(\Bx,t)$ into its
temporal Fourier components $\widehat{\Bs}(\Bx,\Go)$ or Laplace coordinates $\widehat{\Bs}(\Bx,ip)$, and apply the time harmonic equations with time dependence $e^{-i\Go t}$ or equations in the Laplace variable $p$ with time dependence $e^{p t}$, to solve for the response at each frequency $\Go$, or $p$ value, and then integrate these responses.

Given any two fields $\BP_1(\Bx,t)$ and $\BP_2(\Bx,t)$ in this space of fields, we define the inner product of them
to be
\beq (\BP_1,\BP_2)=\int_{\mathbb{R}^4}(\BP_1(\Bx,t),\BP_2(\Bx,t))_{\CT}\,d\Bx dt,
\eeq{innp}
where $(\cdot,\cdot)_{\CT}$ is a suitable inner product on the space $\CT$
such that the projection $\BGG_1$ is selfadjoint with
respect to this inner product, and thus the space $\CE$ onto which
$\BGG_1$ projects is orthogonal to the space $\CJ$ onto which
$\BGG_2=\BI-\BGG_1$ projects. Here we may suppose an experiment is carried out up to a final measurement time $t_0$ and that the fields are zero or small
in the distant past. We may have to add sources or sinks beyond the time $t_0$  to ensure that the fields are also small in the distant future to ensure they
have finite norm.

For examples where $\BL$ has a nontrivial null space
or has ``infinite'' entries,
one may be able to shift $\BL(\Bx)$ and/or its inverse by appropriate
``null-$\BT$ operators'', as mentioned in the Introduction of Part I,
to remove these degeneracies or singularities.

As in the Introduction of Part  I, and
following Section 12.2 of \cite{Milton:2002:TOC} and \cite{Milton:2013:SIG}, 
contact with standard analyses can be made if $\BGG_1(\Bk,\Go)$ (or its blocks, or the corresponding blocks in $\BGG_2(\Bk,\Go)$) has the factorization
\beq \BGG_1(\Bk,\Go)={\BD}(i\Bk,-i\Go)[\BF(\Bk,\Go)]^{-1}{\BD}(i\Bk,-i\Go)^\dagger, \quad \text{where}\quad \BF(\Bk,\Go)={\BD}(i\Bk,-i\Go)^\dagger{\BD}(i\Bk,-i\Go),
\eeq{addd}
in which ${\BD}(i\Bk,-i\Go)$ is a (scalar, vector, tensor, or supertensor) polynomial function of $i\Bk$ and $-i\Go$ and the inverse in \eq{addd} is to be taken on the range of
${\BD}(i\Bk,-i\Go)^\dagger$. In spacetime the action of $\BD$ is given by the differential operator
$\BD(\Grad,\Md/\Md t)$. We can rewrite \eq{ad1} as
\beq \BD(\Grad,\Md/\Md t)^\dagger\BL\BD(\Grad,\Md/\Md t)\BGY=\Bf, \eeq{conv}
and this may be solved for the (possibly multicomponent) potential $\BGY(\Bx,t)$, given a source term $\Bf(\Bx,t)$.
We can identify $\Bf$ with $\BD(\Grad)^\dagger\Bs$, and the potential field $\BGY(\Bx,t)$ may be obtained from $\BE(\Bx,t)$
through their Fourier transforms:
\beq \widehat{\BGY}(\Bk,\Go)=[\BF(i\Bk,-i\Go)]^{-1}{\BD}(i\Bk,-i\Go)^\dagger\widehat{\BE}(\Bk,\Go).
\eeq{conv1}
Conversely, given $\Bf(\Bx,t)$ and a solution $\BGY(\Bx,t)$ and assuming  $\Bs$ is chosen so that $\Bf=\BD(\Grad,\Md/\Md t)^\dagger\Bs$, we get
\beq \BE=\BD(\Grad,\Md/\Md t)\BGY,\quad \BJ=\BL\BE-\Bs,
\eeq{conv2}
Note, as previously, that the source $\Bs$ is not uniquely determined: given a field $\BJ_0$ such that $\BGG_1\BJ_0=0$ we can add it to $\Bs$
and subtract it from $\BJ$, without disturbing $\BE$ or $\BGY$. We can move back and forth between the two different formulations as we please, provided
$\BGG_1(\Bk,\Go)$ (or its blocks, or the corresponding blocks in $\BGG_2(\Bk,\Go)$) have the factorization \eq{addd}.

Clearly \eq{conv} has the solution
\beq \BGY=\underline{\BR}\Bf,\quad\text{where}\quad \underline{\BR}=[\BD(\Grad,\Md/\Md t)^\dagger\BL\BD(\Grad,\Md/\Md t)]^{-1}. 
\eeq{conv3}
As remarked in the Introduction in Part  I, it is not always simpler
to introduce potentials. A spacetime example is a four current $\underline{\Bj}(\underline{\Bx})$ with $\underline{\Bx}=(-t,x_1,x_2,x_3)$
satisfying $\underline{\nabla}\cdot\underline{\Bj}=0$, where $\underline{\nabla}\cdot$ is the spacetime divergence that applied to a 4-vector field
  $\underline{\Bf}(\underline{\Bx})=(f_1(\underline{\Bx}),f_2(\underline{\Bx}),f_3(\underline{\Bx}),f_4(\underline{\Bx}))$ gives
\beq
\underline{\nabla}\Bf=\frac{\Md f_1}{\Md x_1}+\frac{\Md f_2}{\Md x_2} +\frac{\Md f_3}{\Md x_3}-\frac{\Md f_4}{\Md t}.
\eeq{15.14s2}
It has an associated (nonunique)
potential $\BV(\underline{\Bx})$ that is a second order tensor: $j_i=\underline{\Gn}_{ijk\ell}\Md V_{k\ell}/\Md x_j$, where $\underline{\Gn}_{ijk\ell}$
are the elements of the Levi-Civita tensor $\underline{\BGn}$,
that are $1$ or $-1$ according to whether $ijk\ell$ is an even or odd permutation of $1234$. In general, currents can sometimes appear on the right hand
side of the constitutive relation.

As in Part  I, to avoid taking unnecessary transposes, we let $\Div$ or $\underline{\nabla}\cdot$ act on the first index of a field, and the action of $\Grad$ or $\underline{\nabla}$ produces a field, the first index of
which is associated with $\Grad$ or $\underline{\nabla}$.

%%%%%%%%%%%%%%%%%%%%%%%%%%%%%%%%%%%%%%%%%%%%%%%%%%%%%%%%%%%%%%%%%%%%%%%%%%%%%%%%%%%55
\section{Diffusion Equations}
%%%%%%%%%%%%%%%%%%%%%%%%%%%%%%%%%%%%%%%%%%%%%%%%%%%%%%%%%%%%%%%%%%%%%%%%%%%%%%%%
\setcounter{equation}{0}
\labsect{2}
%%%%%%%%%%%%%%%%%%%%%%%%%%%%%%%%%%%%%%%%%%%%%%%%%%%%%%%%%%%%%%%%%%%%%%%%%%%%%%%%%%%55
\subsection{Heat, particle, or light  diffusion with or without convection (advection)}
%%%%%%%%%%%%%%%%%%%%%%%%%%%%%%%%%%%%%%%%%%%%%%%%%%%%%%%%%%%%%%%%%%%%%%%%%%%%%%%%
The heat equation or diffusion equation in a convective fluid having velocity $\Bv(\Bx,t)$  is given by
\beq
\alpha\frac{\partial T}{\partial t} =
\nabla\cdot\BK\nabla T + \Bv\cdot\Grad T-\Bh,
\eeq{A0.68}
where for the heat equation $T(\Bx,t)$ and $h(\Bx,t)$ are the temperature and heat source, $\Ga(\Bx)$ is the product of the heat capacity per unit mass $C_p(\Bx)$
and the mass density $\Gr(\Bx)$ and $\BK(\Bx)$ is the heat conductivity tensor, while for the diffusion equation,  $T(\Bx,t)$ and $h(\Bx,t)$
are the particle concentration and particle source, while for the diffusion equation $\BK(\Bx)$ is the second order diffusivity tensor,
and we may take $\Ga(\Bx)=1$. These equations are important for determining the transport of plumes of pollutants in
the atmosphere or ocean \cite{Stockie:2011:MAD}. 

When manipulated into  the desired form the equation becomes:
\beq
\begin{pmatrix}
\Bq\\
q_t\\
\nabla\cdot\Bq+i\partial q_t/\Md t 
\end{pmatrix}
=\BL
\begin{pmatrix}
\nabla T\\
i\partial T/\Md t \\ T
\end{pmatrix}+\bpm 0 \\ 0 \\ h \epm ,
\eeq{15.12}
where $\Bq(\Bx,t)$ is the heat current or particle current. We found it necessary to introduce the scalar field $q_t(\Bx,t)$.
Now we have
\beqa \BL(\Bx)=\begin{pmatrix}
\BK(\Bx) & 0 & 0\\
0 & 0 & i\tfrac{1}{2}\Ga(\Bx) \\
-\Bv^T(\Bx) & -i\tfrac{1}{2}\Ga(\Bx) &0
\end{pmatrix},\quad \BGG_1(\Bk,\Go)& = & \BG(\Bk,\Go),\quad \text{with}\quad\BG(\Bk,\Go)=\frac{1}{k^2+\Go^2+1}\begin{pmatrix}
\Bk\otimes\Bk & i\Go\Bk & i\Bk \\ 
-i\Go\Bk^T & \Go^2 & \Go \\ -i\Bk^T & \Go & 1
\end{pmatrix}, \nonum
& = &\frac{{\BD(i\Bk,-i\Go)}{\BD(i\Bk,-i\Go)}^\dagger}{k^2+\Go^2+1}\quad\text{with  }\BD(\Grad,\Md/\Md t)=\bpm \Grad \\ i\Md/\Md t \\ 1 \epm,
\eeqa{x12}
where $\Bv(\Bx,t)$ is the velocity of the convecting fluid, and
for the heat equation, $\BK(\Bx)$ is the second order tensor of heat conduction, $\Ga(\Bx)$ the product of the heat capacity per unit mass $C_p(\Bx)$
and the mass density $\Gr(\Bx)$, while for the diffusion equation $\BK(\Bx)$ is the second order diffusivity tensor, and we may take $\Ga(\Bx)=1$.

In the Laplace domain, with a time dependence $e^{pt}$ where $p$ is the Laplace variable the governing equations (until the source is turned off) we have
\beq \bpm \Bq \\ \Div\Bq \epm=\BL \bpm \Grad T \\ T \epm -\bpm 0 \\ \Bh\epm,\quad \BL=\bpm \BK(\Bx,p) & 0\\ -\Bv(\Bx,p) & p\Ga(\Bx,p) \epm,\quad
\BGG_1(\Bk)=\frac{1}{1+k^2}\bpm \Bk \otimes\Bk & i\Bk \\ -i\Bk^T & 1 \epm, \eeq{lp1}
where we have allowed for complex values of $p$ and the real parts of $e^{pt}T$, $\Bq e^{pt}$ and $\Bh e^{pt}$ are now the temperature, heat flux,
velocity field,
and heat source, $k^2=\Bk\cdot\Bk$, and we have let $\Bk$, $\Ga$, and $\Bv$ to depend on $p$ to allow for temporal
relaxation (with one or more relaxation times), and to allow for fluid flows that vary with time. Thus we have obtained an equation
which resembles the acoustic wave equation,
only that $\BL$ for the wave equation is not positive
definite when $\Go$ is real. Neither is the symmetric part of $\BL$ in \eq{x12} positive definite when $\Bv$ is large enough.

For real values of $p$ all the fields are real, and hence $\BL(\Bx,p)$ is too. Provided $\Bv(\Bx,p)$ is small enough that the symmetric part
of $\BL(\Bx,p)$ in \eq{x12} positive definite for all $\Bx$ then we have standard minimization variational principles for which the minimizer
is the solution for a given source $\Bh(\Bx)$ (modulated by $e^{pt}$).
In metamaterials one expects there to be off diagonal couplings in $\BL$ coupling the heat flow with the temperature
and giving an additional heat source linearly related to $\Grad T$.

Light diffusion, in the diffusion approximation, is governed by the same equations but with an additional absorption term. The fluence rate $\GF(\Bx,t)$
plays the role of $T(\Bx,t)$, the current density $\Bj(\Bx,t)$ plays the role of $\Bq(\Bx,t)$ and $\BL(\Bx)$ takes
the form
\beq \BL=\begin{pmatrix}
\BD & 0 & 0\\
0 & 0 & i\tfrac{1}{2}c \\
-\Bv^T & -i\tfrac{1}{2}c & \Gm_a
\end{pmatrix},
\eeq{ld}
where $c(\Bx)$ is the speed of light in the medium, $\BD(\Bx)$ is the diffusion tensor, $\Gm_a(\Bx)$ is the absorption coefficient
and, if we allow for light diffusion in a convective
fluid (with $\BD(\Bx)=D\BI$ and $D$ and $\Gm_a$ independent of $\Bx$), $\Bv$ is the velocity of the convective fluid. Light diffusion
may be used for many biological imaging applications where it is known as diffuse optical imaging (DOI) \cite{Durduran:2010:DOT}.

%%%%%%%%%%%%%%%%%%%%%%%%%%%%%%%%%%%%%%%%%%%%%%%%%%%%%%%%%%%%%%%%%%%%%%%%%%%%%%%%%%%55
\subsection{Linearized reaction diffusion equations
and predator-prey models with diffusion and migration}
%%%%%%%%%%%%%%%%%%%%%%%%%%%%%%%%%%%%%%%%%%%%%%%%%%%%%%%%%%%%%%%%%%%%%%
One of the simplest reaction diffusion equations when two species are present,
associated with the autocatalytic reaction $A+B\rightarrow 2A$, takes the form
\beq \frac{\Md A}{\Md t}=\Div\BD_A\Grad A+kAB+s_A,\quad
\frac{\Md B}{\Md t}=\Div\BD_B\Grad B-kAB+s_B,
\eeq{rde1}
where $A$ and $B$ are the concentrations of the two species;
$\BD_A(\Bx)$ and $\BD_B(\Bx)$ are the diffusion tensors, that we allow to be anisotropic and depend upon $\Bx$ as could be the
case for reaction diffusion in a porous medium; $k$ is the reaction
rate; and $s_A(\Bx)$ and $s_B(\Bx)$ are source terms. When the concentration
$B$ is small relative to $A$ then the term $kAB$ in the equation for
$A$ can be neglected, and $A(\Bx,t)$ satisfies an ordinary diffusion equation
of the same form as in the previous section with $\Bv=0$. Once that has been
solved for $A$ then $B$, which is slaved to it, satisfies
\beq \bpm \Bq_B\\
q_{tB}\\
\nabla\cdot\Bq_B+i\partial q_{tB}/\Md t 
\end{pmatrix}=\BL\bpm \nabla B\\
i\partial B/\Md t \\ B
\end{pmatrix}-\bpm 0 \\ 0 \\ s_B \epm,\quad
\BL=  \bpm  \BD_B(\Bx) & 0 & 0 \\
0 & 0 & \tfrac{1}{2}i  \\
0 & -\tfrac{1}{2}i & kA
\end{pmatrix},\quad  \BGG_1(\Bk,\Go)=\BG(\Bk,\Go).
\eeq{rdeslave}
in which $\BG(\Bk,\Go)$ is given by \eq{x12}. Of course a similar
analysis applies when the concentration
$A$ is small relative to $B$.

More generally, the
equations \eq{rde1} are nonlinear due to the term $kAB$.
Considering the perturbed equations,
where $A$, $B$, $s_A$ and $s_B$ are replaced by
$A+\Ge A'$, $B+ \Ge B'$, $s_A+ \Ge s'_A$ and $s_B+ \Ge s'_B$
we see that, to first order in $\Ge$, the perturbed fields satisfy 
\beq  \frac{\Md A'}{\Md t}=\Div\BD_A\Grad A'+kA'B+kAB'+s_A',\quad
\frac{\Md B'}{\Md t}=\Div\BD_B\Grad B'-kA'B-kAB'+s_B'.
\eeq{rde2}
We can rewrite these as
\beqa
&~& \begin{pmatrix}
\Bq_A'\\
q_{tA}'\\
\nabla\cdot\Bq_A'+i\partial q_{tA}'/\Md t \\
\Bq_B'\\
q_{tB}'\\
\nabla\cdot\Bq_B'+i\partial q_{tB}'/\Md t 
\end{pmatrix}
= \BL
\begin{pmatrix}
\nabla A'\\
i\partial A'/\Md t \\ A' \\
\nabla B'\\
i\partial B'/\Md t \\ B'
\end{pmatrix}-\bpm 0 \\ 0 \\ s'_A \\  0 \\ 0 \\ s'_B \epm, \nonum
&~& \BL(\Bx)  = \begin{pmatrix}
\BD_A(\Bx) & 0 & 0 &  0  & 0 & 0 \\
0 & 0 & \tfrac{1}{2}i  &  0  & 0 & 0 \\
0 & -\tfrac{1}{2}i & -kB &  0  & 0 & -kA \\
 0  & 0 & 0 & \BD_B(\Bx) & 0 & 0 \\
0  & 0 & 0 &0 & 0 & \tfrac{1}{2}i  \\
0  & 0 & kB & 0 & -\tfrac{1}{2}i & kA
\end{pmatrix},\quad \BGG_1(\Bk,\Go) =
\bpm \BG(\Bk,\Go) & 0\\ 0 & \BG(\Bk,\Go)\epm.
\eeqa{rde3}

Predator-prey models are central to mathematical ecology,
and allowing for migration and diffusion take the form \cite{Liu:2010:APP}:
\beq \frac{\Md A}{\Md t}+c_A\Grad A=\Div\BD_A\Grad A+ f(A,B) +s_A,\quad
\frac{\Md B}{\Md t}+c_B\Grad B=\Div\BD_B\Grad B+ g(A,B)+s_B,
\eeq{rde4}
where $A$ and $B$ denote the populations of the two animal species, 
$c_A(\Bx)$ and $c_b(\Bx)$ are migration constants,
$f(A,B)$ and $g(A,B)$, that possibly depend on $\Bx$, include both interaction terms and births
and deaths that are related to population sizes, while $s_A(\Bx,t)$
and $s_B(\Bx,t)$
may represent sources of introduced animals or cullings. Of course if the
terms  $s_A$ and $s_B$ do not depend on time then they can be absorbed
into $f(A,B)$ and $g(A,B)$. However, we choose to keep them separate,
viewing  $s_A$ and $s_B$ as functions
that we can control, and hence perturb.

The
two species could be two competing species, or two species with a
symbiotic relationship and are not necessarily predator and prey. 
Replacing
$A$, $B$, $s_A$ and $s_B$ by $A+\Ge A'$, $B+ \Ge B'$, $s_A+ \Ge s'_A$
and $s_B+ \Ge s'_B$,
we obtain, to first order in $\Ge$, the equations
\beqa \frac{\Md A'}{\Md t}+c_A\Grad A'& = &\Div\BD_A\Grad A'+ f_A(A,B)A'
+f_B(A,B)B'+s_A',\nonum
\frac{\Md B'}{\Md t}+c_B\Grad B'& = &\Div\BD_B\Grad B'+  g_A(A,B)A'
+g_B(A,B)B'+s_B',
\eeqa{rde5}
where $f_A(A,B)$,  $f_B(A,B)$, $g_A(A,B)$, and $g_B(A,B)$ are the partial derivatives:
\beq f_A(A,B)=\frac{\Md f(A,B)}{\Md A},\quad
     f_B(A,B)=\frac{\Md f(A,B)}{\Md B},\quad
     g_A(A,B)=\frac{\Md g(A,B)}{\Md A},\quad
     g_B(A,B)=\frac{\Md g(A,B)}{\Md B}.
     \eeq{rde6}
     The equations take the same form as in \eq{rde3}, but with a new form for
     $\BL(\Bx)$:
     \beq \BL(\Bx)  = \begin{pmatrix}
\BD_A(\Bx) & 0 & 0 &  0  & 0 & 0 \\
0 & 0 & \tfrac{1}{2}i  &  0  & 0 & 0 \\
c_A & -\tfrac{1}{2}i & -f_A(A,B) &  0  & 0 & -f_B(A,B) \\
 0  & 0 & 0 & \BD_B(\Bx) & 0 & 0 \\
0  & 0 & 0 &0 & 0 & \tfrac{1}{2}i  \\
0  & 0 & -g_A(A,B) & c_B & -\tfrac{1}{2}i & -g_B(A,B)
\end{pmatrix},
\eeq{rde7}
and with $\BGG_1(\Bk)$ being the same as before. These more general equations
are also applicable to more complicated reaction-diffusion equations when
two species are present. 
%%%%%%%%%%%%%%%%%%%%%%%%%%%%%%%%%%%%%%%%%%%%%%%%%%%%%%%%%%%%%%%%%%%%%%%%%%%%%%%%%%%%%%%%%%%%%%%%%%%%%%%%%%%%%%%%%%%%%%%%%%%%%%%%%%%%%%%%%%%%%%%%%%%%%%%%%%%%%%%%%%%%%%55
\subsection{The Nernst-Planck equations for the flow of charged particles in a fluid subject to an electric field}
%%%%%%%%%%%%%%%%%%%%%%%%%%%%%%%%%%%%%%%%%%%%%%%%%%%%%%%%%%%%%%%%%%%%%%%%%%%%%%%%
The Nernst-Planck electrodiffusion equations describe the motion of
charged chemical species in a fluid, including biological fluids, when an electric field is present.
These could be cations or anions.
With a fixed electric potential $\Gf(\Bx)$ the
equations take the form \cite{Lu:2010:PNP}:
\beq \frac{\Md \Gr_i}{\Md t}
=\Div [D_i (\Grad \Gr_i+\Gb q_i\Gr_i\Grad\Gf)-\Bv\Gr_i]+h_i,
\eeq{np1}
where $\Gr_i(\Bx,t)$, $i=1,2,\ldots, n$, is the concentration of the $i$-th species (which may be
cations or anions) carrying charge $q_i$, $D_i(\Bx)$ is the spatially
dependent diffusion coefficient,
$h_i(\Bx,t)$ is a source of those particles, and $\Bv(\Bx)$ is the fluid
velocity, assumed to be in steady state flow.
Here $\Gb$ is $1/(k_BT)$ where $k_B$ is the Boltzmann constant, and $T$ is the temperature. We begin by ignoring
the perturbation to $\Gf(\Bx)$ caused by the charged species. We rewrite the equations as
\beq 
\begin{pmatrix}
\Bq_i\\
q_{it}\\
\nabla\cdot\Bq_i+i\partial q_{it}/\Md t 
\end{pmatrix}
=\BL
\begin{pmatrix}
-\nabla \Gr_i\\
-i\partial \Gr_i/\Md t \\ -\Gr_i
\end{pmatrix}+\bpm 0 \\ 0 \\ h_i \epm ,
\eeq{np2}
where $\Bq_i(\Bx,t)$ is the flux of the $i$-species particle current.
As for the other diffusion equations we found it necessary to introduce the
scalar field $q_{it}(\Bx,t)$ (that is not to be confused with the charge $q_i$). Now we have
\beq \BL(\Bx)=\begin{pmatrix}
D_i(\Bx) & 0 & D_i(\Bx)\Gb q_i \Grad\Gf(\Bx)-\Bv(\Bx) \\
0 & 0 & \tfrac{1}{2}i \\
0 & -\tfrac{1}{2}i &0 \end{pmatrix},\quad \BGG_1(\Bk,\Go)=\BG(\Bk,\Go),
\eeq{np3}
where $\BG(\Bk,\Go)$ is defined in \eq{x12}.

If one takes into account the electric field due to the charged
species then the Nernst-Planck equations \eq{np1} need to the supplemented by the dielectric 
(Poisson) equation \cite{Lu:2010:PNP}:
\beq -\Div(\Gve\Grad\Gf)-\sum_{i=1}^n q_i\Gr_i = \Gr_f,
\eeq{np4}
where $\Gr_f$ is the permanent fixed charge distribution, $\Gve(\Bx)$ is the electrical permittivity,
and the electrical potential $\Gf$ is no longer fixed. For simplicity lets just consider the case
where only one species is present $n=1$. The equations are now nonlinear
owing to the term $\Gb q_i \Gr_i\Grad\Gf$ in \eq{np1}. So we replace $\Gf$, $\Gr_i$, $\Bq_{i}$, $q_{it}$, $\Gr_f$, and
$h_i$ with $\Gf+\Ge\Gf'$, $\Gr+\Ge\Gr'$, $\Bq+\Ge\Bq'$, $q_t+\Ge q_t'$, $\Gr_f+\Ge\Gr_f'$, and $h+\Ge h'$,
where we have dropped the subscript $i$ because only one species is present. Then to first order in $\Ge$ the
perturbations satisfy
\beq 
\begin{pmatrix} \Bd' \\ \Div\Bd' \\
\Bq'\\
q_{t}'\\
\nabla\cdot\Bq'+i\partial q_{t}'/\Md t 
\end{pmatrix}
=\BL
\begin{pmatrix} -\Grad\Gf'\\ -\Gf' \\
-\nabla \Gr_i'\\
-i\partial \Gr_i'/\Md t \\ -\Gr_i'
\end{pmatrix}+\bpm 0 \\ \Gr_f' \\ 0 \\ 0 \\ h' \epm ,
\eeq{np5}
and
\beq \BL(\Bx)=\begin{pmatrix} \Gve(\Bx) & 0 & 0 &0 &0 \\ 0 & 0 & 0 &0 & -q \\
 D(\Bx)\Gb q\Gr(\Bx) & 0 &D(\Bx) & 0 & D(\Bx)\Gb q \Grad\Gf(\Bx)-\Bv(\Bx) \\
 0 & 0 &0 & 0 & \tfrac{1}{2}i \\
 0 & 0 &0 & -\tfrac{1}{2}i &0 \end{pmatrix},\quad \BGG_1(\Bk,\Go)=\bpm \Bk\otimes\Bk & i\Bk & 0
\\ -i\Bk^T & 1 & 0 \\ 0 & 0 & \BG(\Bk,\Go)\epm.
\eeq{np6}

\subsection{Small perturbations to diffusion of electrons and holes in semiconductors}
%%%%%%%%%%%%%%%%%%%%%%%%%%%%%%%%%%%%%%%%%%%%%%%%%%%%%%%%%%%%%%%%%%%%%%%%%%%%%%%%
The equations for  diffusion of electrons and holes in semiconductors are \cite{Roosbroeck:1950:TFE}:
\beqa \Div\Bj_n& = & q\Md n/\Md t+qU_n,\quad  \Bj_n=qn\BGm_n\Be+q\BD_n\Grad n, \nonum
      \Div\Bj_p& = & -q\Md p/\Md t+qU_p,\quad \Bj_p=qp\BGm_p\Be+q\BD_p\Grad p,\nonum
            \Bd& = &\BGve\Be,\quad \Div\Bd=q(p-n-N_B),\quad \Be=-\Grad V,
\eeqa{sc0}
where $q$ is the electron charge, $n(\Bx)$ and $p(\Bx)$ are the electron and hole particle densities, $\Bj_n(\Bx)$ and $\Bj_p(\Bx)$ are the associated electrical currents, $\BGm_n(\Bx)$ and $\BGm_p(\Bx)$
are their mobilities (we do not assume the medium is isotropic) connected to their diffusivity tensors $\BD_n(\Bx)$ and $\BD_p(\Bx)$
through the Einstein relations, $\BGm_n=q\BD_n/(k_BT)$ and $\BGm_p=q\BD_p/(k_BT)$ where $k_B$ is Boltzman's constant
and $T(\Bx)$ is the temperature field.
The source terms $U_n$ and $U_p$ are the net generation-recombination rates for electrons and holes, while
$N_B(\Bx)$ represents the net concentration of ionized impurities ($N_B=N_d-N_a$ where $N_d(\Bx)$ and $N_a(\Bx)$ are the concentrations
of ionized donors and acceptors). As the terms $qn\BGm_n\Be$ and $qp\BGm_p\Be(\Bx)$ are nonlinear, we replace
$n$, $p$, $\Bj_n$, $\Bj_p$, $\Bd$, $\Be$, $U_n$, $U_p$, and $N_B$ with $n+\Ge n'$, $p+\Ge p'$, $\Bj_n+\Ge\Bj_n'$, $\Bj_p+\Ge\Bj_p'$,
$\Bd+\Ge\Bd'$, $\Be+\Ge\Be'$, $U_n+\Ge U_n'$, $U_p+\Ge U_p'$ and $N_B+\Ge N_B'$ where $\Ge$ is a small parameter.
Substituting these in \eq{sc0} and collecting the first order terms in $\Ge$ gives the linearized equations:
\beqa \Div\Bj_n'& = & q\Md n'/\Md t+qU_n',\quad  \Bj_n'=qn'\BGm_n\Be+qn\BGm_n\Be'+q\BD_n\Grad n', \nonum
      \Div\Bj_p'& = & -q\Md p'/\Md t+qU_p',\quad \Bj_p'=qp'\BGm_p\Be+qp\BGm_p\Be'+q\BD_p\Grad p',\nonum
            \Bd'& = &\BGve\Be',\quad \Div\Bd'=q(p'-n'-N_B'),\quad \Be'=-\Grad V'.
\eeqa{sc0a}
These can be reexpressed as
\beq \bpm \Bd' \\ \Div\Bd' \\ \Bj_n' \\ j_{nt}' \\ \Div\Bj_n'+i\Md j_{nt}'\Md t \\ \Bj_p' \\ j_{pt}' \\ \Div\Bj_p'+i\Md j_{pt}'\Md t \epm
=\BL \bpm -\Grad V' \\ -V' \\ \Grad n' \\ i\Md n'/\Md t \\ n' \\ \Grad p' \\ i\Md p'/\Md t \\ n' \epm-
\bpm 0 \\ qN_B' \\ 0 \\ 0 \\ qU_n' \\ 0 \\ 0 \\-qU_p' \epm,
\eeq{sc1}
where
\beq \BL=\bpm
\BGve   & 0 & 0      & 0    & 0         & 0       & 0 & 0 \\
      0 & 0 & 0      & 0    & q         & 0       & 0 & -q\\
qn\BGm_n & 0 & q\BD_n & 0    & q\BGm_n\Be & 0       & 0 & 0 \\
      0 & 0 & 0      & 0    & iq/2       & 0       & 0    & 0 \\
      0 & 0 & 0      & -iq/2 & 0         & 0       & 0    & 0 \\
qp\BGm_p & 0 & 0      & 0    & 0         & -q\BD_p & 0    & q\BGm_p\Be \\
      0 & 0 & 0      & 0    &  0        & 0       & 0    & -iq/2 \\
      0 & 0 & 0      & 0    &  0        & 0       & iq/2 & 0\epm,\quad\text{and}\quad
      \BGG_1= \bpm \Bk\otimes\Bk & i\Bk & 0 & 0 \\
                   -i\Bk         & 1    & 0 & 0 \\
                    0            & 0    & \BG(\Bk,\Go) & 0 \\
                    0            & 0    &  0       & \BG(\Bk,\Go)\epm,
\eeq{sc2}     
where the block matrix $\BG(\Bk,\Go)$ is defined in \eq{x12}.  
%%%%%%%%%%%%%%%%%%%%%%%%%%%%%%%%%%%%%%%%%%%%%%%%%%%%%%%%%%%%%%%%%%%%%%%%%%%%%%%%%%%
%%%%%%%%%%%%%%%%%%%%%%%%%%%%%%%%%%%%%%%%%%%%%%%%%%%%%%%%%%%%%%%%%

\subsection{Perturbed spintronic equations}
%%%%%%%%%%%%%%%%%%%%%%%%%%%%%%%%%%%%%%%%%%%%%%%%%%%%%%%%%%%%%%%%%%%%%%%%%%%%%%%%
%%%%%%%%%%%%%%%%%%%%%%%%%%%%%%%%%%%%%%%%%%%%%%%%%%%%%%%%%%%%%%%%%
Spintronics adds a new dimension to electronics, so that one keeps track of the electron spin currents and not just the electrical current. 
Spin polarized currents can be generated
by passing a current through a ferromagnetic material and this is the basis of giant magnetoresistance devices used in the read heads
of magnetic hard drives. 
The two component drift-diffusion spintronic equations, which were originally derived for modeling spin transport in ferromagnetic metals,
describe the motion and diffusion of electrons with spins in a conductor and take the form \cite{Saikin:2005:MSS}:
\beq q\frac{\Md n_{\uparrow(\downarrow)}}{\Md t}=\Div\Bj_{\uparrow(\downarrow)} -
\frac{q}{2\Gt_{sf}}(n_{\uparrow(\downarrow)} -n_{\downarrow(\uparrow)})  + S_{\uparrow(\downarrow)}, \quad \Bj_{\uparrow(\downarrow)} = \BGs_{\uparrow(\downarrow)} \Be +q\BD\Grad n_{\uparrow(\downarrow)},\quad \BGs_{\uparrow(\downarrow)}=q n_{\uparrow(\downarrow)}\BGm,\quad \Be=-\Grad V,
                  \eeq{pse1}
                  where $n_{\uparrow\downarrow)}$ are the density of electrons in the up (down) state and  $\Bj_{\uparrow(\downarrow)}$
                      are the associated currents, $S_{\uparrow(\downarrow)}$ are the associated sources of the spin polarization,
                        $\Gj_{sf}$ is the spin relaxation time,
                      and $\BGm(\Bx)$ is the electron mobility tensor, connected to the electron diffusivity tensor $\BD(\Bx)$
                      via the Einstein relation: $\BGm=q\BD/(k_BT(\Bx))$, where $k_B$ is Boltzman's constant and $T(\Bx)$ is the temperature field.
As the terms $\BGs_{\uparrow(\downarrow)}\Be_{\uparrow(\downarrow)}=qn_{\uparrow(\downarrow)}\BGm\Be_{\uparrow(\downarrow)}$ are nonlinear, we replace
$n_{\uparrow(\downarrow)}$, $\Bj_{\uparrow(\downarrow)}$, $S_{\uparrow(\downarrow)}$, and $V$ with
$n_{\uparrow(\downarrow)}=\Ge n_{\uparrow(\downarrow)}'$, $\Bj_{\uparrow(\downarrow)}+\Ge\Bj_{\uparrow(\downarrow)}$, $S_{\uparrow(\downarrow)}=\Ge S_{\uparrow(\downarrow)}$,
and $V+\Ge V$ where $\Ge$ is a small parameter. Substituting these in \eq{pse1} and collecting the terms that are first order in $\Ge$
gives:
\beq  \bpm 0 \\ \Bj_\uparrow' \\ j_{t\uparrow}' \\ \Div\Bj_\uparrow'+i\Md j_{t\uparrow}'\Md t \\ \Bj_\downarrow' \\ j_{t\downarrow}' \\ \Div\Bj_\downarrow'+i\Md j_{t\downarrow}'\Md t \epm
=\BL \bpm -\Grad V' \\ \Grad n_\uparrow' \\ i\Md n_\uparrow'/\Md t \\ n_\uparrow' \\ \Grad n_\downarrow' \\ i\Md n_\downarrow'/\Md t \\ n_\downarrow' \epm-
\bpm 0 \\ 0 \\ 0 \\ S_{\uparrow}' \\ 0 \\ 0 \\ S_{\downarrow}' \epm,\quad 
\eeq{pse2}
\beq \BL=\bpm
         0        & 0    & 0              & 0            & 0     & 0    & 0 \\
\BGs_{\uparrow}   & q\BD & 0              & q\BGm\Be     & 0     & 0    & 0 \\
      0           & 0    & 0              & -iq/2        & 0     & 0    & 0 \\
      0           & 0    & iq/2           & q/(2\Gj_{sf}) & 0     & 0    & -q/(2\Gj_{sf}) \\
\BGs_{\downarrow} & 0    & 0              & 0            &  q\BD & 0    &  q\BGm\Be \\
      0           & 0    & 0              & 0            &  0    & 0    & -iq/2 \\
      0           & 0    & -q/(2\Gj_{sf}) & 0            &  0    & iq/2 & q/(2\Gj_{sf})\epm,\quad
      \BGG_1= \bpm \Bk\otimes\Bk & 0 & 0 \\
                    0    & \BG(\Bk,\Go) & 0 \\
                    0    &  0       & \BG(\Bk,\Go)\epm.
      \eeq{pse3}

\subsection{Nuclear Magnetic Resonance (NMR) with diffusion}
%%%%%%%%%%%%%%%%%%%%%%%%%%%%%%%%%%%%%%%%%%%%%%%%%%%%%%%%%%%%%%%%%%%%%%%%%%%%%%%%
Here we consider the Bloch-Torrey equations \cite{Torrey:1956:BED} for nuclear magnetic resonance in the presence
of diffusion, neglecting spin-exchange diffusion. Letting $\Bm_0$ denote the equilibrium magnetization,
and $T_1(\Bx)$ and $T_2(\Bx)$ denote the longitudinal and transverse relaxation times, the magnetization $\Bm(\Bt)$ satisfies
\beq \frac{\partial \Bm}{\partial t} = \BC\Bm+\nabla\cdot\BCD\nabla(\Bm-\Bm_0)+(\Gm-\Gm_0)\Bh_0/T_1,
\eeq{nmr1}
where $\BCD(\Bx)$ is the fourth order tensor representing the diffusion in a possibly anisotropic medium, and the matrix $\BC(\Bx)$ is
given by
\beq \BC= \Gg\BGn(\Bh)+\BI/T_2-\frac{(1/T_1-1/T_2)}{h^2}\Bh\otimes\Bh,
%\bpm -1/T_2 & \Gg b_3 & -\Gg b_2 \\ - \Gg b_3 & -1/T_2 &  \Gg b_1 \\  \Gg b_2 & - \Gg b_1 & -1/T_1 \epm,
\eeq{nmr2}
where $\Bh$ is the magnetic field, $h^2=\Bh\cdot\Bh$, $\Gg$ is the gyromagnetic moment, and $\BGn(\Bh)$
is the antisymmetric matrix such that $\BGn(\Bh)\Bm=\Bh\otimes\Bm$. We can recast these equations as
\beq
\begin{pmatrix}
\BQ\\
\Bq_t\\
\nabla\cdot\BM+i\partial \Bq_t/\Md t 
\end{pmatrix}
=\BL
\begin{pmatrix}
\nabla \Bm\\
i\partial \Bm/\Md t \\ \Bm
\end{pmatrix} -\bpm \nabla\Bm_0 \\ 0 \\ -(\Gm-\Gm_0)\Bh/T_1 \epm,
\eeq{nmr3}
which also serves to define the matrix valued flux $\BQ(\Bx,t)$ and the vector field  $\Bq_t(\Bx,t)$.
Now we have
\beq \BL=\begin{pmatrix}
\BCD(\Bx) & 0 & 0\\
0 & 0 & i\tfrac{1}{2}\BI \\
0 & -i\tfrac{1}{2}\BI & -\BC 
\end{pmatrix}, \quad \BGG_1(\Bk,\Go) =  \BG(\Bk,\Go),
\eeq{nmr4}
with $\BG(\Bk,\Go)$ being the same as in \eq{x12} but acting on the first index of $(matrix, vector)$ fields. 
%\quad \BGG_1& = &\frac{1}{k^2+\Go^2+1}\begin{pmatrix}
%\Bk\otimes\Bk & i\Go\Bk & i\Bk \\ 
%-i\Go\Bk^T & \Go^2 & \Go \\ -i\Bk & \Go & 1
%\end{pmatrix} \nonum
%& = &\frac{{\BD(i\Bk,-i\Go)}{\BD(i\Bk,-i\Go)}^\dagger}{k^2+\Go^2+1}\quad\text{with  }\BD(\Grad,\Md/\Md t)=\bpm \Grad \\ i\Md/\Md t \\ 1 \epm,
%\eeqa{x12}
%%%%%%%%%%%%%%%%%%%%%%%%%%%%%%%%%%%%%%%%%%%%%%%%%%%%%%%%%%%%%%%%%%%%%%%%%%%%%%%%%%%55
\section{Radiative transfer and Boltzmann equation in the BGK form}
%%%%%%%%%%%%%%%%%%%%%%%%%%%%%%%%%%%%%%%%%%%%%%%%%%%%%%%%%%%%%%%%%%%%%%%%%%%%%%%%
For radiative transfer, the radiance $L(\Bx,\Bn,t)$ represents the energy flow in the direction $\Bn$ per unit solid angle per unit time, and satisfies the
radiative transfer equation
\beq \frac{1}{c}\frac{\Md L}{\Md t}=-\BV\cdot\Grad L -\BW L +S,
\eeq{rt1}
where $c(\Bx)$ is the speed of light in the medium, $S(\Bx,\Bn,t)$ is the light source,
the gradient just acts on $\Bx$, and the action of $\BV\cdot\Grad$ on $\BL$ produces a scalar field $\Gz(\Bx,\Bn,t)$ given by
\beq \Gz(\Bx,\Bn,t)=\Bn\cdot\Grad L(\Bx,\Bn,t). \eeq{rt2}
Also $\BW$ acts linearly on $L$ to produce a field $\Gx(\Bx,\Bn,t)$ given by
\beq \Gx(\Bx,\Bn,t)=\Gm_t L(\Bx,\Bn,t)-\Gm_s\int_{S_1} P(\Bn',\Bn)L(\Bx,\Bn',t)\,d\Bn', \eeq{rt3}
where the integral is over the sphere $S_1$ of unit radius, $\Gm_t=\Gm_a+\Gm_s$ is the extinction coefficient, $\Gm_s$ the scattering
coefficient, and $\Gm_a$ the absorption coefficient, and $P(\Bn',\Bn)$ is the
probability of light with propagation direction $\Bn'$ being scattered into solid angle $d\Bn'$ around $\Bn'$.
The equation \eq{rt1} can be reexpressed as
\beq
\begin{pmatrix}
\Bj\\
j_t\\
\nabla\cdot\Bj+i\partial j_t/\Md t 
\end{pmatrix}
=\BL
\begin{pmatrix}
\nabla L\\
i\partial L/\Md t \\ L
\end{pmatrix}+\bpm 0 \\ 0 \\ S \epm, \quad \BL=\begin{pmatrix}
0 & 0 & 0\\
0 & 0 & i\tfrac{c}{2} \\
-c\BV^T & -i\tfrac{c}{2} & -\BW
\end{pmatrix},\quad \BGG_1(\Bk,\Go)=\BG(\Bk,\Go),
\eeq{rt4}
where $\BG(\Bk,\Go)$ is defined in \eq{x12}. In the presence of convective flows $\BV$ will also depend on $\Bx$.

For the Boltzmann equation we need to work in phase space so that $(\Bx,t)$ is replaced by $(\Bx,\Bp,t)$ where $\Bp$ is
the momentum. The equation governs the dynamics of $f(\Bx,\Bp,t)$ giving the probability
$f(\Bx,\Bp,t)\,d\Bx\,d\Bp$ of finding a molecule at time $t$ within a volume $d\Bx\,d\Bp$ around the point $(\Bx,\Bp)$
in phase space. The equations take the form
\beq \frac{\Md f}{\Md t}=-\frac{1}{m}\Bp\cdot\Grad f-\BF\cdot\Grad_p f +\BI(f),
\eeq{bzm1}
where $\Grad$ denotes the gradient with respect to $\Bx$, $\Grad_p$ denotes the gradient with respect to $\Bp$,
$m$ is the particle mass, $\BF(\Bx,t)$ is the force field acting on the particles,
and $\BI$ is the interaction term due to collisions. The collision term
is quadratic in $f$, so to obtain a linearized equation one should consider the perturbed equation.
The simplest perturbation is that from a Maxwellian equilibrium distribution function $f_0(\Bx,\Bp)$
at the point $\Bx$ resulting in the Bhatnagar, Gross and Krook (BGK) equations \cite{Bhatnagar:1954:MCP} where
$\BI(f)=\Gv(f_0-f)$ in which $\Gv$ is the molecular collision frequency. These can
be expressed as
\beqa
\begin{pmatrix}
\Bj_x\\ \Bj_p \\
j_t\\
\nabla\cdot\Bj_x+\nabla\cdot\Bj_p+i\partial j_t/\Md t 
\end{pmatrix}
& = & \BL
\begin{pmatrix}
\nabla f\\ \nabla_p f \\
i\partial f/\Md t \\ f
\end{pmatrix}+\bpm 0 \\ 0 \\ 0 \\ vf_0 \epm, \quad \BL=\begin{pmatrix}
  0 & 0 & 0 & 0\\
  0 & 0 & 0 & 0\\
0 & 0 &0 & \tfrac{1}{2}i \\
-\Bp^T/m & -\BF & -\tfrac{1}{2}i & -\Gv
\end{pmatrix},\nonum
\BGG_1(\Bk,\Bk_p,\Go)& = &\frac{1}{k^2+k_p^2+\Go^2+1}\begin{pmatrix}
  \Bk\otimes\Bk & \Bk\otimes\Bk_p & i\Go\Bk & i\Bk \\
   \Bk_p\otimes\Bk & \Bk_p\otimes\Bk_p & i\Go\Bk_p & i\Bk_p \\
   \Go\Bk^T & \Go\Bk_p^T & \Go^2 & \Go \\
   -i\Bk^T & -i\Bk_p^T & \Go & 1
\end{pmatrix}.
\eeqa{bzm2}
in which $\Bk_p$ is the Fourier variable associated with $\Bp$ and $\Grad_p$.

%%%%%%%%%%%%%%%%%%%%%%%%%%%%%%%%%%%%%%%%%%%%%%%%%%%%%%%%%%%%%%%%%%%%%%%%%%%%%%%%%%%%%% 
\section{Perturbed Boussinesq equations for a nonviscous fluid}
\setcounter{equation}{0}
\labsect{Bou}
%%%%%%%%%%%%%%%%%%%%%%%%%%%%%%%%%%%%%%%%%%%%%%%%%%%%%%%%%%%%%%%%%
The Boussinesq equations are for fluid flow where the temperature and density have only
small variation. In the absence of viscosity they are given by
\beq \Gr_0[\Md\Bv/\Md t+(\Bv\cdot\Grad)\Bv]=-\Grad (P+\Gr_0\Bg)-\Ga\Gr_0(T_0-T)\Bg\quad
\Md T/\Md t+(\Bv\cdot\Grad)T=\Div K\Grad T,\quad\Div\Bv=0,
\eeq{Bou1}
where $\Bv(\Bx,t)$, $P(\Bx,t)$, and $T(\Bx,t)$ are the fluid velocity, pressure and
temperature, while $\Ga$ is the thermal expansion coefficient, $K$ is the thermal
conductivity coefficient, $\Gr_0\Bg$ is the gravitational force,
$T_0(\Bx)$ and
$\Gr_0(\Bx)$ are reference temperature and pressure profile. 

As these are nonlinear we consider small perturbations, replacing
$\Bv$, $P$, and $T$ with $\Bv+\Ge\Bv'$, $P+\Ge P'$ and $T+\Ge T'$.
Collecting the first terms in $\Ge$ gives
\beqa -\Grad P'& = &
\Gr_0[\Md\Bv'/\Md t+(\Bv'\cdot\Grad)\Bv+(\Bv\cdot\Grad)\Bv']
-\Ga\Gr_0 T'\Bg+\Grad(\Gr_0\Bg)+\Ga\Gr_0 T_0\Bg,\quad \Div\Bv'=0,\nonum
\Md T'/\Md t& = & \Div K\Grad T'-(\Bv'\cdot\Grad)T-(\Bv\cdot\Grad)T',\quad\Div\Bv'=0.
\eeqa{Bou2}
We rewrite these as
\beq \bpm -P'\BI \\ 0 \\ -\Grad P' \\ \Bq \\ q_t' \\ \Div\Bq'+i\Md q_t'/\Md t \epm
=\BL \bpm \Grad\Bv' \\ i\Md \Bv'/\Md t \\ \Bv' \\ \Grad T' \\ i\Md T'/\Md t \\ T' \epm
-\bpm 0 \\ 0 \\ \Grad(\Gr_b\Bg) -\Ga T_0\Bg \\ 0 \\ 0 \\ 0 \epm,
\eeq{Bou4}
where
\beq \BL=\bpm
  \Gl_h\BGL_h    &   0    &   0    &   0    &   0    &   0    \\
  0    &   0    &   0    &   0    &   0    &   0    \\
  \Bv\cdot    &   0    &  (\Grad\Bv)^T+\Gr_b\BI    &   0   &   0    &   \Ga\Gr_b\Bg   \\
  0    &   0    &   0    &   K\BI   &   0    &   0    \\
  0    &   0    &   0    &   0    &   0    &   i/2    \\
  0    &   0    &   (\Grad T)\cdot    &   \Bv\cdot    &   -i/2    &   0    \epm,
  \quad \BGG_1(\Bk,\Go)=\bpm \BG(\Bk,\Go) & 0 \\ 0 & \BG(\Bk,\Go) \epm,
  \eeq{Bou5}
  in which $\BGL_h$ is the projection onto matrices that are proportional to $\BI$,
  we take the limit $\Gl_h\to\infty$ to ensure $\Div\Bq'=0$. The projection
  $\BG(\Bk,\Go)$ is given by \eq{x12}, with the one in the first block of $\BGG_1(\Bk,\Go)$
  acting on the first index of
the matrix in   $(matrix, vector)$ fields. 
  %%%%%%%%%%%%%%%%%%%%%%%%%%%%%%%%%%%%%%%%%%%%%%%%%%%%%%%%%%%%%%%%%%%%%%%%%%%%%%%%%%%%%% 

\section{Dynamic classical wave equations in stationary media}
\setcounter{equation}{0}
\labsect{3}
%%%%%%%%%%%%%%%%%%%%%%%%%%%%%%%%%%%%%%%%%%%%%%%%%%%%%%%%%%%%%%%%%
When our equations are independent of time $t$, then in a space of three dimensions $\Bx=(x_1,x_2,x_3)$ the associated Fourier space vector
is $\Bk=(k_1,k_2,k_3)$. For equations that depend on time we may take $x_4=t$, so that the associated Fourier space vector
is $\Bk^0=(k_1,k_2,k_3, -\Go)$. In this section, to make the connection with physics easier,
we will write everything in terms of $t$ rather than $x_4$ and $\Go$ and $\Bk$ rather than $\Bk^0$. Since $\Go$ is now a Fourier
coordinate it needs to go in $\BGG_1$ rather than in $\BL(\Bx)$.

We remark that one can recover time harmonic equations at frequency $\Go$ by setting $\Go=\Go_0$ in the expression for $\BGG_1(\Bk,\Go)$
and replacing time derivatives $\Md/\Md t$ with $-i\Go_0$.
Of course $\BGG_1(\Bk)=\BGG_1(\Bk,\Go_0)$ remains a projection operator. However, for many purposes it is best if one filters out
$\Go_0$ from $\BGG_1$, especially if one wants to consider complex $\Go_0$ or time harmonic equations in lossy media with the
equations written so that $\BL(\Bx)$ has a positive semidefinite imaginary part: this is what we did in Part II \cite{Milton:2020:UPLII}.

%%%%%%%%%%%%%%%%%%%%%%%%%%%%%%%%%%%%%%%%%%%%%%%%%%%%%%%%%%%%%%%%%%%%%%%%%%%%%%%%%%%%%%5
\subsection{Dynamic acoustics}
%%%%%%%%%%%%%%%%%%%%%%%%%%%%%%%%%%%%%%%%%%%%%%%%%
Assuming the medium is nondispersive medium with the bulk modulus $\Bk(\Bx)$ and density $\Gr(\Bx)$ being independent
of frequency (and hence ignoring losses) the governing equation for the fluid velocity $\Bv(\Bx)$
and fluid pressure field $P(\Bx)$ is now
\beq \begin{pmatrix}
\partial \Bv/\Md t  \\
\nabla\cdot\Bv
\end{pmatrix}=\BL\begin{pmatrix}
\nabla P\\
-\partial P/\Md t 
\end{pmatrix},
\eeq{15.11a}
We have 
\beqa \BL=\begin{pmatrix}
-\Gr(\Bx)^{-1}\BI & 0 \\
0 & \Gk(\Bx)^{-1}
\end{pmatrix},\quad \BGG_1= \BN(\Bk,\Go)& \equiv &
\frac{{\BD(i\Bk,-i\Go)}{\BD(i\Bk,-i\Go)}^\dagger}{k^2+\Go^2},\quad\text{with}\quad\BD(\Grad,\Md/\Md t)=\bpm \Grad \\ -\Md/\Md t  \epm, \nonum
& = & \frac{1}{k^2+\Go^2}
\begin{pmatrix}
\Bk\otimes\Bk & \Go\Bk \\
\Go\Bk^T & \Go^2
\end{pmatrix}.
\eeqa{x11xx}
%where
%\beq \BN(\Bk,\Go)=
%\frac{1}{k^2+\Go^2}
%\begin{pmatrix}
%\Bk\otimes\Bk & \Go\Bk \\
%\Go\Bk^T & \Go^2
%\end{pmatrix}
%=\frac{{\BD(i\Bk,-i\Go)}{\BD(i\Bk,-i\Go)}^\dagger}{k^2+\Go^2}\quad\text{with}\quad\BD(\Grad,\Md/\Md t)=\bpm \Grad \\ -\Md/\Md t  \epm.
%\eeq{x11xx}
%%%%%%%%%%%%%%%%%%%%%%%%%%%%%%%%%%%%%%%%%%%%%%%%%%%%%%%%%%%%%%%%%%%%%%%%%%%%%5    
 \subsection{Linear  elastodynamic equations}
 Assuming the medium is nondispersive medium with the fourth order elasticity tensor field $\BCC(\Bx)$ and density
 $\Gr(\Bx)$ being independent of frequency (and hence ignoring losses) we have
\beq
\begin{pmatrix}
\partial \BGs/\partial t\\
\nabla\cdot\BGs
\end{pmatrix}
= \BL \begin{pmatrix}
-\nabla \Bv
\\
\partial \Bv/\partial t
\end{pmatrix},
\eeq{15.8}
with
\beq \BL(\Bx)  =  \begin{pmatrix}
-\BCC(\Bx)& 0 \\
0 & \Gr(\Bx)\BI\end{pmatrix},\quad \BGG_1 = \BN(\Bk,\Go),
\eeq{x7a}
where $\BN(\Bk,\Go)$ applied to a (matrix, vector) pair now acts on the first index of the matrix, and on the vector as appropriate.

Here $\Bv=\Md\Bu/\Md t$ is the velocity and  we let $\BL$ do the symmetrizing of $\nabla \Bv$ as this is simpler than working with the strain rate,
$[\nabla \Bv+(\nabla \Bv)^T]/2$.
One may wonder why one has $\partial \BGs/\partial t$ on the left hand side and $\Bv$ on the right hand side
rather than just $\BGs$ and the displacement $\Bu$ when one usually writes the elastic constitutive law in elastodynamics as $\BGs=\BL\BGe$
where $\BGe=[\nabla \Bu+(\nabla \Bu)^T]/2$ is the strain.
However, this is required to get the equations in the desired form \eq{ad1}. An additional justification is that the  elastodynamic equations
should correspond to the acoustic equations when the shear modulus is zero and so that the stress becomes isotropic $\BGs=-P\BI$ where
$P$ is the pressure. In acoustics one does not keep track of the displacement $\Bu$, only the velocity $\Bv=\Md \Bu/\Md t$.

%%%%%%%%%%%%%%%%%%%%%%%%%%%%%%%%%%%%%%%%%%%%%%%%%%%%%%%%%%%%%%%%%%%
\subsection{Dynamic linear thermoelasticity equations}

Assuming the changes in the heat flux $\Bq$
are sufficiently rapid, or the thermal relaxation time $\Gj$ is sufficiently great, so that
\beq \Br\equiv\frac{\Md \Bq}{\Md t}\gg \Bq/\Gj, \eeq{15;1}
then the dynamic linear thermoelasticity equations \cite{Chandrasekharaiah:1986:ACT, Norris:1994:DGF})
take the form
\beq
\begin{pmatrix}
\frac{\partial \BGs}{\partial t}\\
\Div\BGs \\
\Br \\
U \\
\left(\Div\Br+\frac{\Md U}{\Md t}\right)
\end{pmatrix}=\BL\begin{pmatrix}
-\Grad\Bv\\
\frac{\Md \Bv}{\Md t} \\
\Grad\dot\Gt/T_0 \\
\frac{\Md \dot\Gt}{\Md t}/T_0 \\
\dot\Gt/T_0
\end{pmatrix}-\bpm 0 \\ \Bf \\ 0 \\ 0 \\ -\Md \Bh/\Md t \epm.
\eeq{15;2}
Here $\Br$ is the rate of  heat flux $\Bq$ change defined in \eq{15;1}, $U$ is the rate of change of $\Gr S T_0$
where $\Gr$ is the density and $S(\Bx,t)$ is the entropy change, $\Gt(\Bx,t)$ is the change in temperature over
the ambient temperature $T_0$, $\Bf(\Bx,t)$ is the body force and $\Md \Bh/\Md t$ is the rate of change of the
heat source $\Bh(\Bx,t)$. We have
\beqa \BL(\Bx) & = &\begin{pmatrix}
  \BCC & 0 & 0 & -\BGb T_0 &0 \\
  0 & -\Gr\BI & 0 & 0 &0 \\
  0 & 0 & -\BK T_0/\Gj & 0 & 0 \\
  -\BGb T_0 & 0 & 0 & -\Gr c& 0 \\
 0 & 0 & 0 & 0 & 0 
\end{pmatrix},\nonum
\BGG_1 & = &\bpm \BN(\Bk,\Go) &0 \\ 0 & \BY(\Bk,\Go)  \epm,\quad \text{where}\quad
\BY(\Bk,\Go)=\frac{1}{1+k^2+\Go^2}\begin{pmatrix}
\Bk\otimes\Bk & -\Go\Bk & i\Bk \\ 
-\Go\Bk^T & \Go^2 & -i\Go \\
-i\Bk & i\Go & 1
\end{pmatrix}, \nonum &~&
\eeqa{x7b1}
where $c$ is the specific heat per unit mass at constant temperature; the $\BGb(\Bx)=\BGb(\Bx)^T$ is essentially the 
thermal expansion tensor, $\BK$ is the (matrix valued) thermal conductivity tensor. All entries in $\BL(\Bx)$
can depend on $\Bx$ except $T_0$. Note that the structure of $\BL$ forces $\Div\Br+\Md U/\Md t=\Md \Bh/\Md t $. 

%%%%%%%%%%%%%%%%%%%%%%%%%%%%%%%%%%%%%%%%%%%%%%%%%%%%%%%%%%%%%%%%%%%%%%%%%%%%%%%%%%%
%%%%%%%%%%%%%%%%%%%%%%%%%%%%%%%%%%%%%%%%%%%%%%%%%%%%%%%%%%%%%%%%%%%%%%%%%%%%%%%%%%%%%%%%%%%%%%%
\subsection{Dynamic  piezoelectric equations} 
In nondispersive media these take the form \cite{Auld:1973:AWF, Norris:1994:DGF}:
\beq
\begin{pmatrix}
\partial \BGs/\Md t \\
\nabla\cdot\BGs \\
\Md \Bd/\Md t
\end{pmatrix}
=\BL\begin{pmatrix}
-\frac{1}{2}\left[\nabla \Bv+\nabla\Bv^T\right]
\\
\partial \Bv/\Md t \\
\Md \Be/\Md t
\end{pmatrix}, \quad \Div\Bd=0,\quad \Be=-\Grad V.
\eeq{15.10}
Here of course, because we are not using the full electromagnetic equations, we are assuming the time variations of the fields are sufficiently
slow that we can apply the quasistatic approximation for the relation between $\Be(\Bx,t)$ and $\Bd(\Bx,t)$, while keeping the 
dynamic equations for the relations between $\BGs$ and $\Bv$, that then couple with $\Be$ and $\Bd$.
\beq \BL(\Bx)=\begin{pmatrix}
-\BCC(\Bx) & 0 & -\BCA(\Bx)\\ 
0 & \Gr(\Bx)\BI & 0 \\
-\BCA^T(\Bx) & 0 & \BGve(\Bx)
\end{pmatrix},\quad \BGG_1=\bpm \BN(\Bk,\Go) & 0 \\ 0 & \Bk\otimes\Bk/k^2\epm,
\eeq{x7b2}
where now $\BCC(\Bx)$ is the elasticity tensor when the electric field is zero, and $\BCA(\Bx)$ is some third order tensor
piezoelectric coupling term, and $\BN(\Bk,\Go)$ defined in \eq{x11xx} acts on the first index of the matrix in $(matrix, vector)$ fields.
%%%%%%%%%%%%%%%%%%%%%%%%%%%%%%%%%%%%%%%%%%%%%%%%%%%%%%%%
\subsection{Biot's Dynamic  poroelastic equations} 

In nondispersive media these are given by  \cite{Biot:1962:MDA, Norris:1994:DGF}
\beq
\begin{pmatrix}
\partial \BGs/\Md t \\
\nabla\cdot\BGs \\
-\partial P/\Md t \\ 
-\Grad P
\end{pmatrix}
=\BL\begin{pmatrix}
-\nabla \Bv \\
\partial \Bv/\Md t \\ -\Div\Bw \\
\Md \Bw/\Md t
\end{pmatrix},
\eeq{15.10biot}
where $P$ is the fluid pressure, $\Bv=\Md \Bu/\Md t$ is the velocity in the solid component, the time derivative of its displacement,
$\Bw$ is the velocity of the fluid component.
\beq \BL(\Bx)=\begin{pmatrix}
-\BCC(\Bx) & 0 & \BM & 0\\ 
0 & \Gr(\Bx) & 0 & \Gr_f \\
\BM & 0 & M & 0 \\
0 & \Gr_f & 0 & m
\end{pmatrix},\quad \BGG_1=\bpm \BN(\Bk,\Go) & 0 \\ 0 & \BN(\Bk,\Go) \epm,
\eeq{x7b3}
where the $\BM(\Bx)$ are elastic moduli coupling the solid and fluid rates of deformation; $\Gr(\Bx)$ is the density of the solid phase; $\Gr_f$ the density of the fluid
phase which, as it is transported, is assumed to be constant; $M$ is the coefficient relating the pressure change to a volumetric change in fluid content ($\Div\Bw$)
at isovolumetric strain $\Div\Bu=0)$; and $m$ is the inertial resistance of the pore space to an
inviscid fluid.
We are assuming the frequency is high enough that this term dominates the permeability of the pore space, including
the viscous resistance of the fluid pore space that is governed in the time harmonic equations by the imaginary part of the permeability $\Bk(\Bx)$ \cite{Johnson:1987:TDP}.
Otherwise $m$ needs to be replaced by a convolution operator in time \cite{Biot:1962:MDA, Norris:1994:DGF}. The medium is then dispersive.
Note that the first $\BN(\Bk)$ in $\BGG_1(\Bk)$ acts on the first index of the matrix in $(matrix, vector)$ fields
while the second $\BN(\Bk)$ acts on $(vector, scalar)$ fields.

%%%%%%%%%%%%%%%%%%%%%%%%%%%%%%%%%%%%%%%%%%%%%%%%%%%%%%%%%%%%%%%%%%%%%%%%%%%%%%%%%%%%%%%%%
\subsection{Full dynamic electromagnetic equations} 

We make the assumption that the medium is nondispersive, i.e. that the moduli entering the constitutive relation do not depend on frequency (otherwise
we would typically need convolutions in time). 
This can be a reasonable assumption if $\BGm(\Bx)$ and $\BGve(\Bx)$ are real and
$\BGm(\Bx)\geq\Gm_0\BI$ and $\BGve(\Bx)\geq\Gve_0\BI$ for all $\Bx$, where $\Gm_0$ and $\Gve_0$ are the permeability and permittivity
of the vacuum. (Ultimately, at extremely high frequency, both $\BGm(\Bx)/\Gm_0$ and $\BGve(\Bx)/\Gve_0$ should approach $1$ but could remain
relatively constant over the frequency window of interest that is dictated by the time variations in the fields). By contrast, if at a given frequency
$\BGm(\Bx)/\Gm_0$ or $\BGve(\Bx)/\Gve_0$ is less than 1 then the medium is necessarily dispersive, with derivatives with respect to frequency: see
the optimal bounds (6)  in \cite{Milton:1997:FFR}.
\beq \begin{pmatrix}
-\Bh\\
\Bd
\end{pmatrix}
= \BL\begin{pmatrix} \Bb\\ \Be\end{pmatrix}-\bpm \Bs_1 \\ \Bs_2 \epm,\quad \begin{pmatrix} \Bb\\ \Be\end{pmatrix}=\begin{pmatrix}\Curl & 0\\
  -\frac{\Md}{\Md t} & -\Grad\end{pmatrix}\bpm \BGF \\ V \epm,
\quad \begin{pmatrix}\Curl & \frac{\Md}{\Md t}\\
                         0 & \Div \end{pmatrix}\begin{pmatrix}-\Bh\\ \Bd \end{pmatrix}=0.
\eeq{15.14}
To see how  $\Bs_1$ and $\Bs_2$  are related to the free current $\Bj_f$ and free charge density $\Gr_f$ we note that
\beq \Gr_f=\Div\Bd,\quad \Bj_f=\Curl\Bh-\frac{\Md \Bd}{\Md t}=\underline{\nabla}\cdot\bpm \BGn(\Bh) \\ \Bd^T \epm,
  \eeq{15.14s1}
  where $\underline{\nabla}\cdot$ defined by \eq{15.14s2} acts on the first index of a $4\times 3$ matrix valued field,
  and $\BGn$ has indices $\Gn_{ijk}$ that are $1$ or $-1$ according to whether $ijk$ is an even or odd permutation
  of $1,2,3$. Thus $\Curl\Bh=\Div\BGn(\Bh)$, where $\BGn(\Bh)$ is an antisymmetric matrix. So integrating the first equation in \eq{15.14s1} with respect to time,
  and the second with respect to spacetime (column by column) gives the (nonuniquely defined) sources $\Bs_2$ and $\Bs_1$. Now we have
\beqa \BL(\Bx) =  \begin{pmatrix}
-[\BGm(\Bx)]^{-1} & 0\\
0 & \BGve(\Bx)
\end{pmatrix},\quad
\BGG_1&  = & \frac{\BD(i\Bk,-i\Go)\BD(i\Bk,-i\Go)}{k^2+\Go^2}, \quad\text{with  }\quad\BD(\Grad,\Md/\Md t)=\bpm \BGn(\Grad) & 0 \\ -\BI\Md/\Md t & -\Grad \epm, \nonum
& = & 
\frac{1}{k^2+\Go^2}\bpm k^2\BI-\Bk\otimes\Bk & \Go\BGn(\Bk) \\ -\Go\BGn(\Bk) &  \Go^2\BI+\Bk\otimes\Bk  \epm.
\eeqa{x14}
\section{Dynamic quantum wave equations with stationary potentials}
\setcounter{equation}{0}
\labsect{3.7}
%%%%%%%%%%%%%%%%%%%%%%%%%%%%%%%%%%%%%%%%%%%%%%%%%%%%%%%%%%%%%%%%%%%%%%%%%%%%%%%%%%    
 \subsection{Full dynamic multielectron Schr{\"o}dinger equation}
 If the sources are varying in time, then that can cause transitions between different energy states and generally one
 will need the full dynamic equation:
\beq
\begin{pmatrix}
\Bq_x\\
q_t\\
\nabla\cdot \Bq_x+\partial q_t/\Md t 
\end{pmatrix}
=\BL \begin{pmatrix}
\nabla\Gy\\
\partial \Gy/\Md t \\
\Gy
\end{pmatrix},
\eeq{15.19}
where $\Gy(\Bx,t)=\Gy(\Bx_1,\Bx_2,\ldots,\Bx_N,t)$ is the time dependent wavefunction at time $t$, $\Bq_x(\Bx,t)=\Bq_x(\Bx_1,\Bx_2,\ldots,\Bx_N,t)$
is a $3N$-component vector field, and $\Bq_t(\Bx,t)=q_t(\Bx_1,\Bx_2,\ldots,\Bx_N,t)$ is a scalar field. We have
 \beq \BL(\Bx)=\begin{pmatrix}
-\BA & 0 & 0\\
0 & 0 & -\frac{i \hbar}{2}\\
0 & \frac{i\hbar}{2} & -V(\Bx)
\end{pmatrix}, \quad \BGG_1(\Bk,\Go)=\BS(\Bk,\Go),
\eeq{x19a}
where $V(\Bx)$ is the potential, $\BA$ in the simplest approximation is $\hbar^2\BI/(2m)$ in which $m$ is the mass of the electron, but it may take other forms to take into account the reduced mass of the
electron, or mass polarization terms, due to the motion of the atomic nuclei
mass of the electron,
\beqa \BS(\Bk,\Go)&\equiv&\frac{{\BD(i\Bk,-i\Go)}{\BD(i\Bk,-i\Go)}^\dagger}{k^2+\Go^2+1},\quad\text{with}\quad \BD(\Grad,\Md/\Md t)=\bpm \Grad\\ \Md/\Md t \\ 1 \epm, \nonum
& = & \frac{1}{k^2+\Go^2+1}\begin{pmatrix}
\Bk\otimes\Bk & -\Go\Bk & ik \\ 
-\Go\Bk^T & \Go^2 & -i\Go \\ -ik & i\Go & 1
\end{pmatrix}.
\eeqa{x19}
As written, the full dynamic multielectron Schr{\"o}dinger equation \eq{15.19} has no source term. However, as is standard
knowledge, source terms can arise when one perturbs the potential by $\Ge V'(\Bx,t)$. Specifically, they arise in the equation satisfied by
$\Gy'(\Bx,t)$ when $\Gy(\Bx,t)$ is replaced by $\Gy(\Bx,t)+\Ge\Gy'(\Bx,t)$ and  $V(\Bx)$ is replaced by $V(\Bx)+\Ge V'(\Bx,t)$
(and other dependent fields are correspondingly perturbed) in the equations \eq{15.19} and terms in first order in $\Ge$ are equated. The analysis
is a special case of the analysis in Section 13 of Part I and we will not repeat it.

It is well known that the Schr{\"o}dinger equation corresponds to the diffusion equation with imaginary time. Comparing these equations with the diffusion equations \eq{15.12} and \eq{x12}
it is perhaps more natural to say the converse: that the diffusion equation corresponds to the  Schr{\"o}dinger equation with imaginary time.

%%%%%%%%%%%%%%%%%%%%%%%%%%%%%%%%%%%%%%%%%%%%%%%%%%%%%%%%%%%%%%%%%%%%%%%%%%%%%%%%%%%%5555
 \subsection{Full dynamic single electron Schr{\"o}dinger equation in a magnetic field}

It takes the same form as \eq{15.19} but now, as we have only a single electron, $\Bx=\Bx_1$ and $\Gy(\Bx,t)=\Gy(\Bx_1,t)$,
$\Bq_x(\Bx)=\Bq_x(\Bx_1)$ is a 3-component field, $q_t(\Bx)=q_t(\Bx_1,t)$ remains a scalar field, and $\BL$ gets replaced by that in \eq{x20}. Taking $\hbar=1$, we have
\beq
\BL(\Bx)=\begin{pmatrix}
\frac{-\BI}{2m} & 0 & \frac{\ii e \BGF(\Bx)}{2m}\\
0 & 0 & -\frac{\ii}{2}\\
\frac{-\ii e \BGF(\Bx)}{2m} & +\frac{\ii}{2} & -q V(\Bx)
\end{pmatrix}, \quad  \BGG_1(\Bk,\Go)=\BS(\Bk,\Go),
\eeq{x20}
where $\BGF(\Bx)$ is the time independent magnetic potential,
with $\Bb=\Curl\BGF$ the magnetic induction,
$V(\Bx)$ the time independent electric potential,
$q$ the charge on the electron, and $m$ its mass.
%%%%%%%%%%%%%%%%%%%%%%%%%%%%%%%%%%%%%%%%%%%%%%%%%%%%%%%%%%%%%%%%%%%%%%%%%%%%%%%%%%%%5555
\subsection{Perturbations of two fluid models for superfluid liquids and gases }
Tisza in 1938 realized that the features of superfluid Helium were indicative of
two fluid hydrodynamics associated with a Bose-Einstein condensate. He predicted second sound which allows heat to
be transferred as wave and as a consequence superfluid Helium is a better thermal conductor than any other known material.
The effect of second sound is most noticeable when one cools liquid Helium below the superfluid
$\Gl$-transition point where the boiling Helium becomes instantly quiet because conduction of heat through waves is
far more efficient than conduction of heat through diffusion. More recently
second sound has proved to be an important tool in the study of quantum turbulence.
Landau in 1941 introduced a suitable two fluid mathematical
model that is the basis for our treatment here. For an excellent review of these developments see \cite{Donnelly:2009:TFT} and references therein.
Landau's  model has also proved applicable to calculating the frequencies of the low lying Landau two fluid hydrodynamic modes in a
trapped Fermi superfluid gas at unitarity \cite{Taylor:2008:VTT, Donnelly:2009:TFT}.

The superfluid is considered to be a mixture of two fluids. As we are considering the linearized equations,
the normal component is taken to have density $\Gr_n+\Ge\Gr_n'$ and velocity $\Ge\Bv_n$ and
the superfluid component is taken to have density $\Gr_s+\Ge\Gr_s'$ and velocity $\Ge\Bv_s$, where $\Ge$ is a scaling parameter that is small.
We now follow Taylor, Hu, Liu, and Griffin \cite{Taylor:2008:VTT},
who express the equations in terms of the two fluid displacements $\Ge\Bu_n$ and $\Ge\Bu_s$ whose time derivatives
are the velocities $\Ge\Bv_n$ and $\Ge\Bn_s$. 
The entropy density $S+\Ge S'$  is just carried by the normal fluid, so
conservation of entropy, and  mass imply that to first order in $\Ge$:
\beq 0=S'+\Div(S_0\Bu_n),\quad
0= \Gr'+\Div(\Gr_n \Bu_n+\Gr_s \Bu_s)=\Gr'-(\Gr_n/S)S'+\Div(\Gr_s \Bu_s),
\eeq{sf0}
where the mass density is $\Gr+\Ge\Gr'=\Gr+\Ge(\Gr'_n+\Gr_s')$. Then the two fluid equations are given by
\beq \bpm (\Gr_{n}/S)S' \\ \Bp_n \\ \Gr'-(\Gr_{n}/S)S' \\ \Bp_s \epm = \BL
\bpm \Div\Bu_n \\ -\Md \Bu_n/\Md t \\ \Div\Bu_s \\ -\Md \Bu_s/\Md t \epm, \quad \Grad[(\Gr_{n}/S)S']+\Md \Bp_n/\Md t=0,\quad
\Grad[\Gr'-(\Gr_{n}/S)S']+\Md \Bp_s/\Md t=0,
\eeq{sf4}
where $\Bp_n$ and $\Bp_s$ are the normal and superfluid momentum,  
\beq \BL=\bpm c_1 & 0 & c_2 & 0  \\ 0 & -\Gr_n & 0 & 0 \\ c_2 & 0 & c_3 & 0 \\ 0 & 0 & 0 & -\Gr_s \epm,
\quad \BGG_1= \bpm \BN(\Bk,\Go) & 0 \\ 0 & \BN(\Bk,\Go) \epm,
\eeq{sf5}
and $\BN(\Bk,\Go)$ is given by \eq{x11xx}. The terms appearing in $\BL$ can be identified through the quadratic form associated with $\BL$ that is
given in equation (38) in \cite{Taylor:2008:VTT}:
\beq c_1=-\Gr_s^2\left(\frac{\Md \Gm}{\Md \Gr}\right)_S,\quad c_2=-S\Gr_s\left(\frac{\Md T}{\Md \Gr}\right)_S,\quad
c_3=S^2\left(\frac{\Md T}{\Md S}\right)_\Gr+2S\Gr_n\left(\frac{\Md T}{\Md \Gr}\right)_S,
\eeq{sf6}
in which the temperature $T$ and chemical potential per unit mass $\Gm$ are functions of the independent variables
$\Gr$ and $S$ and are given in terms of the energy density $U$ and trapping potential $V_{\rm{ext}}(\Bx)$ through
\beq T=\left(\frac{\Md U}{\Md S}\right)_\Gr,\quad \Gm=\left(\frac{\Md U}{\Md \Gr}\right)_S+ V_{\rm{ext}}.
\eeq{sf7}

%%%%%%%%%%%%%%%%%%%%%%%%%%%%%%%%%%%%%%%%%%%%%%%%%%%%%%%%%%%%%%%%%%%%%%%%%%%%%%%%%%%%%%%%%%%%%%%%%%%%%%%%%%%%%%%%%%%%%%%%%%%%%%%%%% 
\section{Coupling terms in dynamic wave equations when the material moduli depend on time}
%%%%%%%%%%%%%%%%%%%%%%%%%%%%%%%%%%%%%%%%%%%%%%%%%%%%%%%%%%%%%%%%%%%%%%%%%%%%%%%%%%%%%%%%%%%%%%%%%%%%%%%%%%%%%%%%%%%%%%%%%%%%%%%%%%
\setcounter{equation}{0}
\labsect{4}

For the acoustic, elastic, and electrodynamic equations in a medium with properties that do not depend on time,
the tensor $\BL(\Bx)$ has a block diagonal structure. When the material is dynamic or with time dependent fluctuations
in its material dependent properties, it is tempting to just let the moduli depend on time. Thus, for example,
in the elastodynamic equations one expects them to have the same form but with a time dependent density $\Gr(\Bx,t)$
and a time dependent elasticity tensor $\BCC(\Bx,t)$. However, if this dynamic medium can be considered to be a
normal elastic substance that has time dependent fluctuations in the moduli due to its movement with a velocity
field $\Bw(\Bx)$ then one quickly realizes that additional coupling terms necessarily must occur in $\BL(\Bx)$.
This has been recognized for a long time \cite{Kaplan:1930:FCE, Post:1962:FSE} (see also Section 3.4.2 of \cite{Serdyukov:2001:EBAM},
the book \cite{Lurie:2007:IMT} and Section 3 of \cite{Milton:2006:CEM}) and here we review this.

To begin with, let us examine the acoustic equations \eq{15.11a} and \eq{x11xx}. Suppose we are in three (spatial) dimensions. Let $x_4=-t$, and set
\beq \underline{\nabla}=\begin{pmatrix}\nabla\\ \frac{\Md}{\Md x_4}\\\end{pmatrix}=\begin{pmatrix} \nabla\\ -\frac{\Md}{\Md t}
\end{pmatrix},\quad
j_i=\frac{\Md v_i}{\Md t},\quad{\rm for}~i=1,\ldots 3,\quad j_4=\Div\Bv.
\eeq{A0.33a}
Then, with $\underline{\Bx}=(-t,x_1,x_2,x_3)$, the equations \eq{15.11a} can be rewritten as
\beq \underline{\nabla}\cdot\Bj(\underline{\Bx})=0,\quad \Bj(\underline{\Bx})=\BL\underline{\nabla}P(\underline{\Bx}),
\eeq{A0.34b}
so they look like a ``spacetime'' version of the conductivity equations,%
\index{conductivity equations}
with $\Bj(\underline{\Bx})$ satisfying the
conservation law $\underline{\nabla}\cdot\Bj(\underline{\Bx})=0$.
Of course, there are important differences: most
notably the matrix entering the constitutive law is real but not positive definite, as it should be for a wave equation.

Now, by direct analogy with the transformation of the conductivity
equations under affine transformations (see, for example, Section 8.3 of \cite{Milton:2002:TOC}), under the Galilean
transformation
\beq \underline{\Bx}'=\BA(\Bx,t)\underline{\Bx},\quad{\rm with}~ \BA(\Bx,t)=\begin{pmatrix} 1 & \Bw(\Bx,t)\\ 0 & \BI\end{pmatrix},
\eeq{A0.34c}
where $\underline{\Bx}'=(x'_1,x'_2,x'_3,-t')$, $\underline{\Bx}=(x_1,x_2,x_3,-t)$, and
$\Bw(\Bx,t)$ is the velocity of the moving frame of reference, the equations transform to
\beq \underline{\nabla}'\cdot\Bj'(\underline{\Bx}')=0,\quad \Bj'(\underline{\Bx}')=\BL'(\underline{\Bx}')\underline{\nabla}P'(\underline{\Bx}'),
\eeq{A0.34d}
in which
\beqa \BL'& = & \BA\BL\BA^T=\begin{pmatrix}
-\Gr(\Bx)^{-1}\BI+\Gk(\Bx)^{-1}\Bw\Bw^T &  \Gk(\Bx)^{-1}\Bw\\
\Gk(\Bx)^{-1}\Bw^T & \Gk(\Bx)^{-1}
\end{pmatrix},
  \nonum
          \Bj'(\underline{\Bx}')& = & \BA\Bj(\underline{\Bx}),
          \quad P'(\underline{\Bx}')=P(\underline{\Bx}),\quad \underline{\nabla}'P'(\underline{\Bx}')=(\BA^T)^{-1}\underline{\nabla}P(\underline{\Bx}).
\eeqa{A0.34e}
Thus in a moving fluid in Eulerian coordinates the tensor $\BL$ will not be diagonal, and the density term will not
be isotropic.

If the material parameters in Lagrangian coordinates have some dependence on time then $\Gr$ and $\Gk$ in \eq{A0.34e} will
depend both on space and time, $\Gr=\Gr(\Bx,t)$ and $\Gk=\Gk(\Bx,t)$. In general, though, there is no reason to expect that
the form of $\BL(\Bx,t)$ should be restricted to the form of \eq{A0.34e}. If one accepts an off diagonal coupling term,
as one must, then why cannot it too have a general dependence on space and time? Indeed, in a medium with a general time dependence
$\BL(\Bx,t)$ it seems that Lagrangian coordinates should lose their meaning altogether and that $\BL(\Bx,t)$ should at least take
the general form
\beq \BL(\Bx,t)=\begin{pmatrix} \BGr(\Bx,t) & \Bf(\Bx,t) \\ \Bf(\Bx,t)^T & \Gk(\Bx,t)^{-1} \epm,
  \eeq{-1}
  where the mass density term $\BGr(\Bx,t)$ may be anisotropic and the vector fields $\Bd(\Bx,t)$ and $\Bf(\Bx,t)$ induce coupling.
  Then the equations will also be invariant under more general spacetime transformations.

 Similar conclusions apply to the elastodynamic equations \eq{15.8} and \eq{x7a}. Following Section 1.10 of \cite{Milton:2016:ETC} set
\beq \underline{\nabla}=\begin{pmatrix}\nabla\\\frac{\Md}{\Md x_4}\end{pmatrix}=\begin{pmatrix}\nabla\\-\frac{\Md}{\Md t}
\end{pmatrix},\quad
J_{ik}=-\frac{\Md \Gs_{ik}}{\Md t},\quad{\rm for}~i,k=1,2,3,\quad J_{4k}=-\{\Div\BGs\}_k,
\eeq{A0.36a}
which defines the $4\times 3$ matrix valued field $\BJ(\Bx)$. Then noting that, due to the symmetry of $\BCC$,
we can replace $[\nabla\Bv+(\nabla\Bv)^T]/2$
by $\nabla\Bv$ in \eq{15.8}, the equations take the form
\beq \underline{\nabla}\cdot\BJ=0,\quad \BJ=\BL\underline{\nabla}\Bv.
\eeq{A0.36b}
Under the Galilean transformation
\eq{A0.34c} the new fields based on the transformation laws
\beq \BJ'(\underline{\Bx}')=  \BA(\underline{\Bx})\BJ(\underline{\Bx}), \quad
\underline{\nabla}'\Bv(\underline{\Bx}')=[\BA(\underline{\Bx})^T]^{-1}\underline{\nabla}\Bv(\underline{\Bx}),
\eeq{A0.36ba}
become
\beqa
\begin{pmatrix}
\frac{\partial \BGs'}{\partial t'}\\
\nabla'\cdot\BGs'
\end{pmatrix}
& = &\begin{pmatrix} \BCI & \Bw\BI \\
0 & \BI \end{pmatrix}
\begin{pmatrix}
\frac{\partial \BGs}{\partial t}\\
\nabla\cdot\BGs
\end{pmatrix}=\begin{pmatrix}
\frac{\partial \BGs}{\partial t}+\Bw(\Div\BGs)^T\\
\nabla\cdot\BGs
\end{pmatrix},\nonum
\begin{pmatrix}
-\nabla' \Bv'
\\
\frac{\partial \Bv'}{\partial t'}
\end{pmatrix} & = &\begin{pmatrix} \BCI & 0 \\
\BI\Bw^T & \BI \end{pmatrix}^{-1}\begin{pmatrix}
-\nabla \Bv
\\
\frac{\partial \Bv}{\partial t}
\end{pmatrix}
 =  \begin{pmatrix}
-\nabla \Bv
\\
\frac{\partial \Bv}{\partial t}+\Bw^T\nabla \Bv
\end{pmatrix},
\eeqa{A0.36c}
in which $\BCI$ is the fourth order identity tensor, $\BI$ is the second order identity tensor,
and $\Bw\BI$ is a third order tensor with elements $w_i\Gd_{jk}$ (in which $\Gd_{jk}$ is one if $j=k$ and zero otherwise). These
fields are now linked by the new constitutive tensor
\beqa \BL'(\underline{\Bx}')& = & \begin{pmatrix} \BCI & \Bw\BI \\
0 & \BI \end{pmatrix}\BL(\Bx)\begin{pmatrix} \BCI & 0 \\
\BI\Bw^T & \BI \end{pmatrix} \nonum
& = &
\begin{pmatrix}-\BCC(\Bx)+\Bw\BGr(\Bx)\Bw^T & \Bw\BGr(\Bx)\\
\BGr(\Bx)\Bw^T & \BGr(\Bx) \end{pmatrix}.
\eeqa{A0.36d}
Under more general transformations, such as a Galilean transformation followed by a rotation, the fields in \eq{A0.36c} will be also
multiplied on the right by rotation matrices (and the transformation matrix $\BA$ would also need to be adjusted). However
if it is a pure Galilean transformation then there is no multiplication on the right by a rotation matrix: note that since $\Bu(\Bx)$
represents an assumed small difference between the position of a particle in the undeformed state (now moving) and the position of the
particle in the deformed state, $\Bv$ being the difference between two velocities is itself invariant under a Galilean transformation.

By analogy with \eq{-1} we would expect that $\BL(\Bx,t)$ may take
the general form
\beq \BL(\Bx,t)=\begin{pmatrix} \BCC(\Bx,t) & \BF(\Bx,t) \\ \BF(\Bx,t)^T & \BGr(\Bx,t) \epm,
  \eeq{-10}
that is invariant under spacetime transformations \cite{Milton:2006:CEM}. 
  With these more general forms of $\BL(\Bx, t)$ energy is not conserved and the solutions may blow up in time. One can see this most easily if one looks
  at spacetime microstructures and the simplest of this is a ``spacetime laminate'' where the material moduli are only a function of $\Bk\cdot\Bx-\Go t$.
  The easiest way to generate such waves is to propogate through a medium a large amplitude ``pumping wave'' which due to nonlinearities effectively
  creates a ``spacetime laminate'' that can affect a small amplitude wave propaging through the medium, resulting in parametric amplification \cite{Cassedy:1963:DRT}. Conventional media that are periodic is space, but with moduli that do not depend on time,
  sometimes exhibit band gaps in $\Go$ where over a range of frequencies no waves propogate \cite{John:1987:SLP, Yablonovitch:1987:ISE}.
  They reflect waves at these frequencies: for fixed real $\Go$ in the band gap the dispersion relation $\Go=f(\Bk)$ is only satisfied with complex values of $\Bk$ corresponding
  to wave amplitudes that decay exponentially as one moves inside the material. Materials that are periodic in both space and time, can now have bandgaps in
  frequency and the dispersion relation $g(\Bk,\Go)=0$ with real values of $\Bk$ may only have solutions with complex values of $\Go$ that can correspond
  to waves that grow exponentially in time. In the presence of dissipation there is a fine balance, generally the waves either increase exponentially with
  time or decrease exponentially with time, but if the material parameters are carefully tuned there can be waves that propogate, neither increasing exponentially
  nor dying out \cite{Torrent:2018:LCT}. Such a careful tuning may not be necessary if there are nonlinearities.
  Sometimes there are no bandgaps in $\Bk$. An example occurs within the special class of ``spacetime'' periodic microstructures
  that generate ``field patterns'' where the characteristic lines form a pattern commensurate with the material, but not neccessarily with the same unit cell of
  periodicity \cite{Milton:2017:FP}. For these structures the waves propagate with no  exponential increase or decay in amplitude \cite{Mattei:2017:FPW}:
  they are an example of "$PT$-symmetry" \cite{Bender:2007:MSN}. 
  An interesting question is whether such propagating waves always occur when there are no bandgaps in any ``timelike'' direction in $(\Bk,\Go)$ space.
  
  Following the treatise of Lurie on the subject \cite{Lurie:2007:IMT}, the study of waves in spacetime microstructures has drawn increasing attention.
  See, for example, the recent overview by Caloz and Deck-L\'eger \cite{Caloz:2020:SMG,Caloz:2020:SMT}. 
  Besides field patterns, there are a number of other interesting effects in ``spacetime microstructure'' including: oneway wave propagation \cite{Lurie:1997:EPS} and other
  nonreciprocal phenonema; quantum like behavior \cite{Couder:2006:SPD};
  energy accumulation originating in a sort of ``spacetime'' shock wave \cite{Lurie:2006:WPE, Lurie:2009:MAW}; and the generation of a sort of ``effective magnetic
  field'' for photons \cite{Fang:2012:REM}. Time like interfaces also allow time reversal imaging \cite{Bacot:2016:TRH}. 

\section*{Acknowledgements}
GWM thanks the National Science Foundation for support through grant DMS-1814854,
Like Parts I and II \cite{Milton:2020:UPLI, Milton:2020:UPLII}, this Part is also largely based on the books \cite{Milton:2002:TOC, Milton:2016:ETC}
and again I thank those (cited in the acknowledgements of Part I) who helped stimulate that work and who provided feedback on the drafts of those books. In particular,
Nelson Beebe is thanked for all the work he did on preparing the books for publication and for updating the associated bibtex entries.
%%%%%%%%%%%%%%%%%%%%%%%%%%%%%%%%%%%%%%%%%%%%%%%%%%%%%%%%%%%%%%%%%%%%%%%%%%%%%%%%%%%%%%%%%%
%\bibliographystyle{plain}
%\bibliography{/home/milton/tcbook,/home/milton/newref}
%%%%%%%%%%%%%%%%%%%%%%%%%%%%%%%%%%%%%%%%%%%%%%%%%%%%%%%%%%%%%%%%%%%%%%%%%
\ifx \bblindex \undefined \def \bblindex #1{} \fi\ifx \bbljournal \undefined
  \def \bbljournal #1{{\em #1}\index{#1@{\em #1}}} \fi\ifx \bblnumber
  \undefined \def \bblnumber #1{{\bf #1}} \fi\ifx \bblvolume \undefined \def
  \bblvolume #1{{\bf #1}} \fi\ifx \noopsort \undefined \def \noopsort #1{}
  \fi\ifx \bblindex \undefined \def \bblindex #1{} \fi\ifx \bbljournal
  \undefined \def \bbljournal #1{{\em #1}\index{#1@{\em #1}}} \fi\ifx
  \bblnumber \undefined \def \bblnumber #1{{\bf #1}} \fi\ifx \bblvolume
  \undefined \def \bblvolume #1{{\bf #1}} \fi\ifx \noopsort \undefined \def
  \noopsort #1{} \fi

\end{document}